# A Low Complexity Scheme for Entanglement Distributor Buses

Majid Ghojavand

Physics Department, Sharif University of Technology, P.O. Box 11365-9161, Tehran, Iran

## ABSTRACT

For technological purposes and theoretical curiosity, it is very interesting to have a building block that produces a considerable amount of entanglement between on-demand sites through a simple control of a few sites. Here, we consider permanently-coupled spin networks and study entanglement generation between qubit pairs to find low-complexity structures capable of generating considerable entanglement between various qubit pairs. We find that in axially symmetric networks the generated entanglement between some qubit pairs is rather larger than generic networks. We show that in uniformly-coupled spin rings each pair can be considerably entangled through controlling suitable vertices. To set the location of controlling-vertices, we observe that the symmetry has to be broken for a definite time. To achieve this, a magnetic flux can be applied to break symmetry via Aharonov-Bohm effect. Such a set up can serve as an efficient entanglement distributor bus in which each vertex-pair can be efficiently entangled through exciting only one fixed vertex and controlling the evolution time. The low-complexity of this scheme makes it attractive for use in nanoscale quantum information processors.

# Introduction

Entanglement generation and distribution are two central problems in performing many quantum information tasks such as perfect-state transferring[1], parallel processing[2][3], and dense coding[4]. In interacting spin systems, the natural interactions between different nodes can be used for entangling various vertex-pairs. Taking into account their low level of complexity, these systems can be considered in the entanglement provision in realistic quantum modules. To be able to entangle each qubit pair of the system efficiently, the quantum state of the system must be capable of being dynamically controlled. This is due to entanglement monogamy inequality, that is at each time the sum of the pair-wise entanglements of a vertex with the other vertices can never exceed one e-bit (entanglement of a Bell state) [5]. Therefore, to entangle a vertex with another one quite well, the entanglement of the vertex with the other vertices, if any, must be suppressed by changing the quantum state of the system. More precisely speaking, the quantum state of the system had to be controllable to be fitted into some proper quantum states, in each of which some qubits are highly entangled. In one of the controlling mechanisms, the targeted qubit-pair, i.e. those qubits which we intend to entangle, are externally controlled [6][7][8]. However, external interactions produce more noise and more constituents are needed for their implementation as well. In order to change the quantum state properly, one can exploit the natural evolution of the quantum state in permanently coupled spin systems, as opposed to external control. Here, we show that in such a system, the quantum state of the system can pass through all the proper quantum states, which are highly entangled qubit-pair states. Such a module can serve as a low complexity entanglement distributor bus to entangle any qubit-pair of the system selectively.

We consider a general structure of spin networks in which a few qubits are initialized and analytically compute entanglement between each qubit pair in different times. Using this general analytic relation, we propose a low-complexity scheme capable of efficient entangling of any pair of system qubits.

The organization of the paper is as follows: In section I we introduce the notion of *efficient entanglement distributor bus* and explain its importance. In section II we compute entanglement between different vertex-pairs for general structures. Using the general relation obtained in section II, we search for special spin systems that can act as efficient entanglement distributor buses in sub-sections II-A and II-B. This paper is closed by a summary in section III.

## I.  ENTANGLEMENT DISTRIBUTOR BUS

In this section, we explain the notion of "entanglement distributor bus". A "bus" is conceptually distinct from a single "channel". In computer architecture, a bus is a subsystem that transfers data between computer components inside a computer or between computers. Unlike a point-to-point connection, a bus can logically connect several peripherals over the same set of wires. So using buses instead of single wires plays a critical role in the reduction of the system complexities and the number of components in order to achieve more compact and reliable modules. Complexity reduction is of greater importance in quantum processors than in classical processors because of the destructive role of decoherence and error accumulation [10]. So, performing defined quantum actions with low-complexity modules is a central issue for the implementation feasibility of complex quantum procedures. Consequently, using entanglement distributor buses plays a critical role in the realization of more complex quantum actions.

By an *efficient entanglement distributor bus*, we mean a system *i)* that can efficiently entangle each vertex-pair of the system and *ii)* whose structural and operational complexity is as low as possible. To further reduce *system's complexity*, we must try to pursue the following strategies in designing quantum modules : *a)* favoring global rather than regional control *b)* controlling only a few vertices *c)* applying local rather than non-local operations *d)* initializing similarly for different targeting (no need for coding) and *e)* employing un-modulated structures instead of engineered ones.

## II.  THEORY

Consider $N$ interacting qubits labeled by 1, 2, …, N. Our problem is to entangle selectively qubit pairs by controlling the initial quantum state of some of the qubits, called controlling-qubits, labeled by $c_1, c_2, ..., c_{n_0}$ where $c_j$'s are integers that $1 \leq c_j \leq N$ and $n_0$ is the number of controlling-qubits. This means that in the initialization process only the quantum state of controlling-qubits may be adjusted and the state of the other qubits remains intact. We assume qubits interact via a ferromagnetic Hamiltonian, e.g. Heisenberg, XX, or more generally XXZ. Thus

$$\left[H, \sum_{i=1}^{N}\sigma_i^z\right] = 0 \qquad (1)$$

The presence of several real nanoscale examples of such systems, such as Quantum-Dot arrays [11] and atoms trapped in optical-lattices [12], motivates us to bring them into our consideration. The ground states of such systems are so that all of the spins are parallel and downward, so $|\lambda_0\rangle = |\underline{0}\rangle = \bigotimes_{j=1}^{N}|0\rangle_i = |00...0\rangle$ where by 0 and 1 we mean spins $-\frac{\hbar}{2}$ and $\frac{\hbar}{2}$ respectively. This state can be easily prepared by extreme cooling of the system in the presence of a uniform weak magnetic field. The singly-excited subspace is defined as $K = span\{|k\rangle : 1 \leq k \leq N\}$ where $|k\rangle = \left|00...\underbrace{1}_{k-th\,site}...00\right\rangle$. We assume that the initial state is a superposition of the ground state $|\lambda_0\rangle$ and a singly-excited state of controlling-qubits

$$|\Psi(0)\rangle = Sin\theta\, e^{i\varphi}|\underline{0}\rangle + Cos\theta \sum_{i=1}^{n_0} \alpha_i |c_i\rangle \qquad (2)$$

Due to equation (1), the number of excitations of quantum states is preserved, thus the evolved quantum state will still be a superposition of the ground state and a singly-excited state, so

$$|\Psi(t)\rangle = Sin\theta\, e^{i\varphi}|\underline{0}\rangle + Cos\theta \sum_{i=1}^{n_0} \alpha_i \left(\sum_{j=1}^{N} f_{jc_i}(t)|j\rangle\right) \qquad (3)$$

where

$$f_{ji}(t) := \langle j|e^{-iHt/\hbar}|i\rangle = \sum_{k=1}^{N}\langle j|\lambda_k\rangle\langle\lambda_k|i\rangle e^{-i\omega_k t} \tag{4}$$

Here $f_{ji}(t)$ is the *transition probability amplitude* of state $|i\rangle$ at $t_0 = 0$ to state $|j\rangle$ at time $t$ and $|\lambda_k\rangle$ is the $k$th singly-excited eigenstate and $\omega_k := \dfrac{E_k}{\hbar}$ is its eigenfrequency.

Defining a new variable

$$A_j(t) := \sum_{i=1}^{n_0} \alpha_i f_{jc_i}(t) \tag{5}$$

we have

$$|\Psi(t)\rangle = Sin\theta\, e^{i\varphi}|0\rangle + Cos\theta \sum_{j=1}^{N} A_j(t)|j\rangle \tag{6}$$

where $A_j(t)$, called the *excitation-amplitude*, is the probability amplitude for the excitation at the vertex $j$ when the initial state is purely singly-excited ($\theta = 0$).

Inserting (4) into (5) we find

$$A_j(t) = \sum_{k=1}^{N} B_{j,k}\, e^{-i(\omega_k t + \phi_{j,k})} \tag{7}$$

where $B_{j,k}$ and $\phi_{j,k}$ are complex constants. It is also evident that

$$\sum_{j=1}^{N}|f_{ij}(t)|^2 = \sum_{j=1}^{N}|A_j(t)|^2 = 1 \tag{8}$$

In order to compute the entanglement generated between two qubits $m$ and $n$, we should first determine the reduced density matrix of these two qubits, $\rho_{mn}(t)$. Then we use concurrence of $\rho_{mn}(t)$ as the measure of entanglement [13]. The concurrence of $\rho$ is defined as $C = \max\{0, \tau_1 - \tau_2 - \tau_3 - \tau_4\}$, where $\tau_i$'s are the square root of eigenvalues of the non-Hermitian matrix $R = \rho(\sigma_y \otimes \sigma_y)\rho^*(\sigma_y \otimes \sigma_y)$ in decreasing order and $\sigma_y$ is the Pauli spin-flip operator. For the case of (6) the entanglement of two qubits $m$ and $n$ can be calculated as (see appendix A)

$$C_{mn}(t) = 2Cos^2(\theta)|A_m(t)A_n(t)| \tag{9}$$

Since the excitation-amplitudes are independent of initialization parameters $\theta$ and $\varphi$, the first step of entanglement maximization quickly reads as $\theta_{opt} = 0$, so

$$|\Psi(0)\rangle_{opt} = \sum_{i=1}^{n_0} \alpha_i |c_i\rangle \tag{10}$$

And the evolved state (6) becomes

$$|\Psi(t)\rangle = \sum_{j=1}^{N} A_j(t)|j\rangle \tag{11}$$

Henceforth, our mission will be to maximize the product of each pair of the excitation-amplitudes as the target function

$$C_{mn}(t) = 2|A_m(t) A_n(t)| \tag{12}$$

Therefore the maximum obtainable entanglement depends generally on the set of couplings $(\omega_k)$, initialization parameters $(\alpha_i)$ and also targeted pair $(m, n)$ for a spin system. In generic systems, the optimal initialization parameters depend on the targeted pair, consequently coding is necessary. Recall that the necessity of coding increases the complexity of the process.

The excitation-amplitudes fluctuate in time due to the dispersion of initial wave function and its self-interference. Moreover, due to (7), each excitation-amplitude evolves by N dominating harmonic terms (N eigen-frequencies and relative amplitudes and phases). In systems other than those with deliberately engineered couplings, each harmonic term contributes in each excitation-amplitude $A_j(t)$ with different weight through unrelated eigen-frequencies $\omega_k$, amplitudes $B_{j,k}$ and phases $\phi_{j,k}$. Therefore, in generic systems, various excitation-amplitudes are uncorrelated in time and accordingly their absolute maximums occur at different times. However, according to (12), in order to achieve a high level of entanglement between two vertices their relevant excitation-amplitudes should adopt their absolute maximums at the same time. So, in generic systems, we encounter the following typical limitation in achieving a rather large amount of entanglement between vertex-pairs

$$C_{mn\_\max} = 2\left|A_m(t)A_n(t)\right|_{\max} << 2\left|A_m(t)\right|_{\max}\left|A_n(t)\right|_{\max} \tag{13}$$

To see in what extent the poor correlation of excitation-amplitudes limits the obtainable entanglement, we consider the case of Fig 3-a which is a uniformly coupled spin ring initialized as $|\Psi(0)\rangle = |1\rangle$. Due to symmetric conditions of this example, the excitation-amplitudes of a few vertex-pairs are completely correlated in time. So, we are able to compare the entanglement of uncorrelated pairs (e.g. the pair of 4th and 8th qubits) with the entanglement of correlated ones (e.g. the pair of 4th and 14th qubits). The details are illustrated in fig 1.

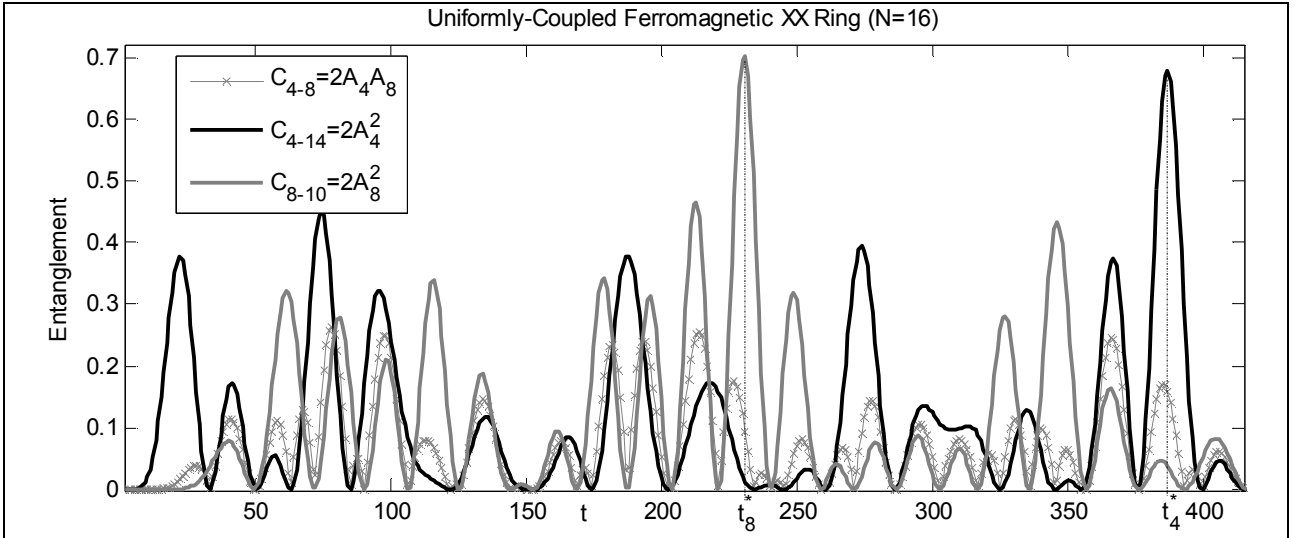

Fig (1): comparison between the entanglement of uncorrelated vertex-pairs($C_{4-8}$) with the entanglement of correlated vertices ($C_{4-14}$ and $C_{8-10}$) in the case of Fig 3-a: As it is seen, the absolute maximums of generic excitation-amplitudes $|A_4(t)|$ and $|A_8(t)|$ do not occur at the same time ($t_4^* \neq t_8^*$), so that when the maximum value of one of them occurs, the other term is relatively negligible. This considerably diminishes their product, i.e. $C_{4-8} = 2|A_4(t)A_8(t)|$

In short, in generic situations the lack of correlation between different $A_i$'s limits entanglement, and for entanglement maximization, the initial state must be coded for the target pair. In the next sub-section, we show that in some symmetric systems not only the strong inequality (13) converts to equality for some pairs, but also the optimal initial state is the same for maximum entanglement of different pairs (no coding is needed). Accordingly, we would be able to devise an efficient entanglement distributor bus.

## A) Axially symmetric systems

To convert inequality (13) to equality for some $A_i$'s, we set up conditions in which the excitation propagates symmetrically in the graph around an axis. This can be done by using an axially symmetric structure which is initialized symmetrically (see Fig. 2). This strategy partitions the graph into some strata $\{\Gamma_i : 0 \leq i \leq d\}$, in each of which the vertices has equal excitation-amplitude all the time. We show the number of elements of $i$th stratum by $n_i$. In axially symmetric structures, various vertices of a stratum interact in the same way with the other strata. Since initialization must also be symmetrical, we identify one stratum $\Gamma_0$ as the control region and symmetrically initialize its vertices as

$$|\Psi(0)\rangle_{opt} = \frac{1}{\sqrt{n_0}} \sum_{k \in \Gamma_0} |k\rangle \tag{14}$$

Associating a unit vector $|\phi_i\rangle := \frac{1}{\sqrt{n_i}} \sum_{k \in \Gamma_i} |k\rangle$ to each stratum (we call this $i$th stratum state), the initial state (14) and the evolved state (11) reform to

$$|\Psi(0)\rangle_{opt} = |\phi_0\rangle$$

$$|\Psi(t)\rangle = \sum_{i=0}^{d} F_i(t) |\phi_i\rangle$$

where

$$F_i(t) := \sqrt{n_i} A_j(t) \tag{15}$$

and $j$ is the index of one of the vertices of $i$th stratum. In fact $F_i(t)$ is the amplitude of excitation over $i$th stratum. Considering (8), we find $\sum_{i=0}^{d} |F_i|^2 = 1$ where the summation is over all strata. As an alternative definition, *axially symmetric structures* are systems in which the evolution of a symmetrical initial quantum state is restricted to $span\{|\phi_i\rangle : 0 \leq i \leq d\}$.

Hereafter, we call those vertex-pairs that have equivalent excitation-amplitude at all times as *targetable-pairs* because a larger amount of entanglement is generated between those pairs at the optimal time.

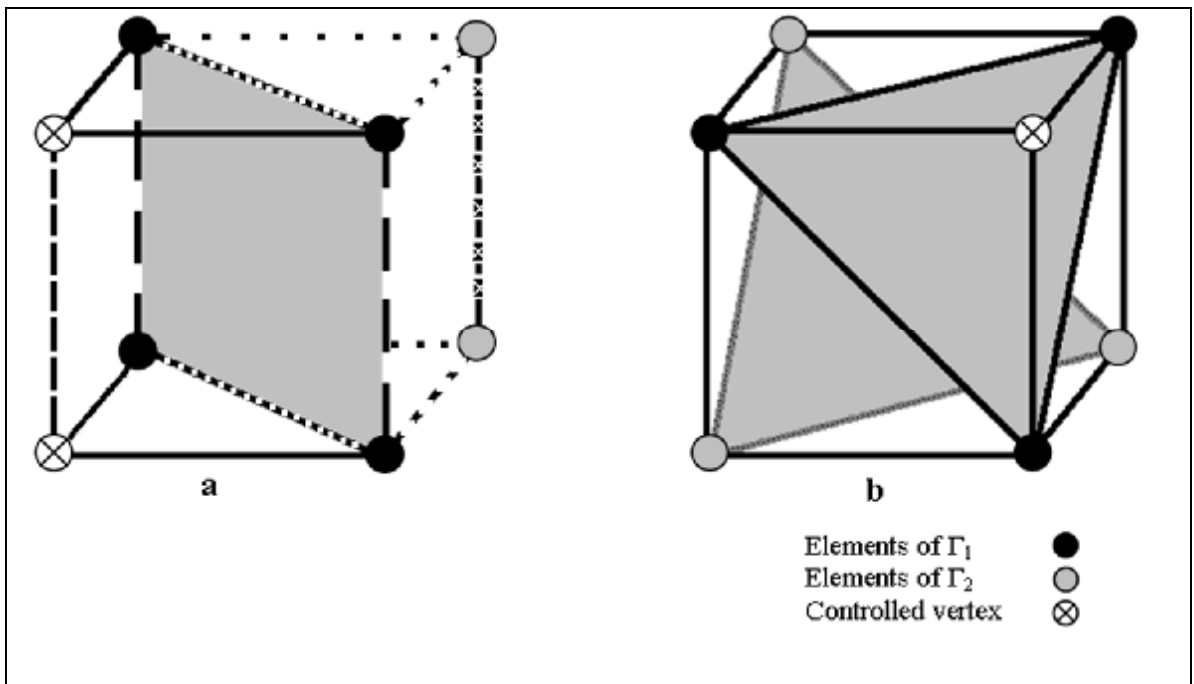

Fig (2): Examples of symmetric systems that excitation-amplitude of vertices located in the same strata can be equalized by symmetric initializing of elements of a stratum. In a graph, the symmetric propagation of excitation may be possible in different ways by symmetrically initializing various vertices as depicted in (a) and (b). The interaction couplings can be various as considered in (a).

Using (15) and (12) the entanglement of a targetable-pair is obtained as

$$C_{ii\_\max} = \frac{2}{n_i}|F_i|^2_{\max} \tag{16}$$

For vertex-pairs located in the different strata (non-targetable pair), the result is

$$C_{ij\_\max} = \frac{2}{\sqrt{n_i n_j}}|F_i(t)F_j(t)|_{\max} \tag{17}$$

where the typical limitation (13) still remains valid (see Fig. 1).

To simplify the procedure, the control region can be chosen among the strata with lowest population. Particularly, the procedure is much simplified if $n_0 = 1$ since, in this case, the initial state (14) can be prepared without any nonlocal operation; otherwise it will be a Bell-state or a W-state. Eq. (14) implies that the initial state is the same for targeting different targetable pairs in axially symmetric systems whereas in a generic system the controlling-vertices must be coded for the location of the intended pair to create the maximum value of entanglement.

Since entanglement of the pairs located in the same strata is proportional to $n_i^{-1}$ (see (16)), it is preferred to use one-dimensional axially symmetric graphs because of their low populated strata, i.e. $n_i \leq 2$. On the other hand, the implementation of one-dimensional systems is generally simpler than other graph structures (In fact, there are several well-known nanoscale candidates for them [14][15][16]). So, hereafter we consider symmetric one-dimensional spin systems to devise efficient entanglement distributor buses.

**B) One Dimensional Entanglement Distributor Buses**

Because of their advantages for entanglement distribution, in this sub-section, we will consider one-dimensional systems to design an efficient entanglement distributor bus. Following our proposed strategy for saturating inequality (13) in one-dimensional systems, we require that the map of interaction couplings enjoy mirror symmetry w.r.t. a line so that the strength of each coupling and its dual (mirror-picture) is equal (see Fig. 3). In this case, each stratum contains two or one vertices. If one of the vertices (e.g. $c$) is located on the symmetry line, its stratum does not contain any other vertex and it is the best candidate for controlling (see Fig 3-a). So, we will use

$$|\Psi(0)\rangle_{opt} = |c\rangle \tag{18}$$

; Otherwise, the related initial state (14) is a Bell state (see Fig. 3-b)

$$|\Psi(0)\rangle_{opt} = \frac{1}{\sqrt{2}}\left(|c\rangle + |\bar{c}\rangle\right) \tag{19}$$

where $\bar{c}$ is the index of the vertex that is the mirror-picture of vertex $c$ (e.g. in Fig. 3-b $\bar{1} \equiv 16$). The initial state (18) is much simpler than (19) to be prepared. In practice, the initialization (18) is simply done by applying a local magnetic field in the Z direction at the controlling-vertex $c$ while cooling. However, for initialization (19) a nonlocal operation is needed.

For those systems that have one symmetry line, the ratio of targetable pairs to total pairs is $\frac{1}{N}$. However, the ideal situation is to reach systems in which each vertex-pair is targetable. At first glance, very high symmetric structures, such as uniformly-coupled spin rings, seem to be the solution because in such systems vertex-pairs are the mirror-picture of each other w.r.t. one of the N symmetry lines of the system. But for targeting different pairs related to various symmetry lines a different vertex (or vertex-pair) must be controlled. To target all pairs, every vertex must be controllable.

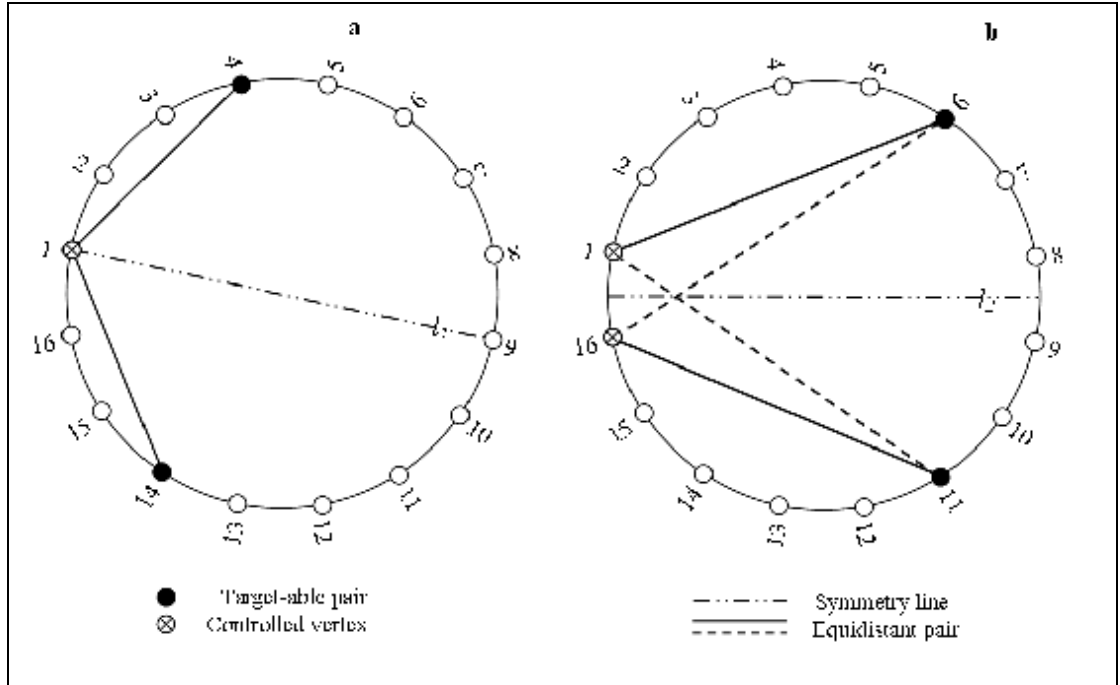

Fig (3): The presence of each symmetry line causes the relative mirror symmetric vertices to be targetable. In a uniformly-coupled even ring, half of the symmetry lines pass from a pair of antipodal vertices so that the relative initialized state is a local excited state(3-a); while the other symmetry lines don't cross any vertex (3-b). In the latter, the analogous initialized state has the form (19).

Recall that we desire to target all pairs by controlling minimum number of vertices. To achieve this, we can transfer the excitation to another vertex by a global controlling without using any local agitation. Such a process is a type of perfect state transfer (P.S.T.), i.e. $|f_{ij}|=1$. However, while the symmetric conditions govern the system, P.S.T. is impossible. To verify this impossibility, let $\alpha_1 = 1 \text{ and } \alpha_2 = ... = \alpha_{n_0} = 0$, then due to (8) we find $|f_{ij}| \leq \frac{1}{\sqrt{n_j}}$. This result shows that a fairly complete state transfer from the vertex $i$ to one of the vertices of non-single strata $j$ can never happen. So, in homogeneous spin rings, aside from the special case of $j = N/2 + i$ for even N, we have $|f_{ij}| \leq \frac{1}{\sqrt{2}}$. Thus, the operated global field must break such symmetry in a finite period of time.

We can use a magnetic flux that passes through the ring to break rotational symmetry of the Hamiltonian via Aharonov-Bohm effect (see appendix 2). The Hamiltonian is

$$H = -\frac{J}{2}\sum_{k=1}^{N}\left[(e^{i\Phi}\sigma_k^+ . \sigma_{k+1}^- + H.c.)\right] + \Delta\sum_{k=1}^{N}\sigma_k^z . \sigma_{k+1}^z \quad (20)$$

Where $\Phi$ is the magnetic flux. The singly-excited eigenvalues and eigenvectors of (20) can be obtained as [9]

$$E_m = 4J\cos[(2\pi/N)(m+f)] - J(N-4)$$

$$|\lambda_m\rangle = \sum_{k=1}^{N} e^{i(2\pi/N)(m+f)k}|k\rangle$$

where $f := \Phi/2\pi$. Accordingly

$$f_{ij}(t) = e^{i(2\pi f/N)(j-i)}\sum_{k=1}^{N}e^{i[-E_k t/\mathbf{h} + 2\pi k(j-i)/N]}$$

$$|f_{ij}(t)| = \left|\sum_{k=1}^{N}e^{i[-E_k t/\mathbf{h} + \varphi_k]}\right| \quad (21)$$

where $\varphi_k := 2\pi k(j-i)/N$. To study the effect of the magnetic flux in symmetric propagation of wave-function, we compare $|f_{ij}(t)|$ with $|f_{i\bar{j}}(t)|$ where $\bar{j} = 2i - j$ is the mirror-picture of $j$ w.r.t. $i$. The result is

$$|f_{i\bar{j}}(t)| = \left|\sum_{k=1}^{N} e^{i[-E_k t/\hbar + \varphi_{N-k}]}\right| \qquad (22)$$

For $f = Nq$ where $q \in Z$, we have $E_k = E_{N-k}$; so, due to (21) and (22) $|f_{ij}(t)| = |f_{i\bar{j}}(t)|$ while for $q = 0$, the symmetric propagation of the wave function in the pre-discussed situation is corroborated. For $f \neq Nq$, we have $E_k \neq E_{N-k}$. So, we will not wait for a symmetric situation while the relevant simulations confirm the non-symmetric propagation of the wave function. The numerical studies show that high fidelity transfer to each desired destination vertex can occur in this way [9]. Particularly, in homogeneous spin rings consisting of 5 or 7 vertices, near-perfect transfer of excitation (greater than 0.999) from each vertex to another is possible [9]. The fact that the total number of vertices in these two systems is odd is helpful because, in such rings, each symmetry line passed through one of the vertices (see Fig 4). Thus, the optimal initial state is of the simple type(18). In short, to entangle each desired pair in such systems, we should excite one fixed vertex and control the magnetic flux for transferring the excitation on the relevant symmetry line and then turn off the field and allow the system to evolve naturally for an optimal time.

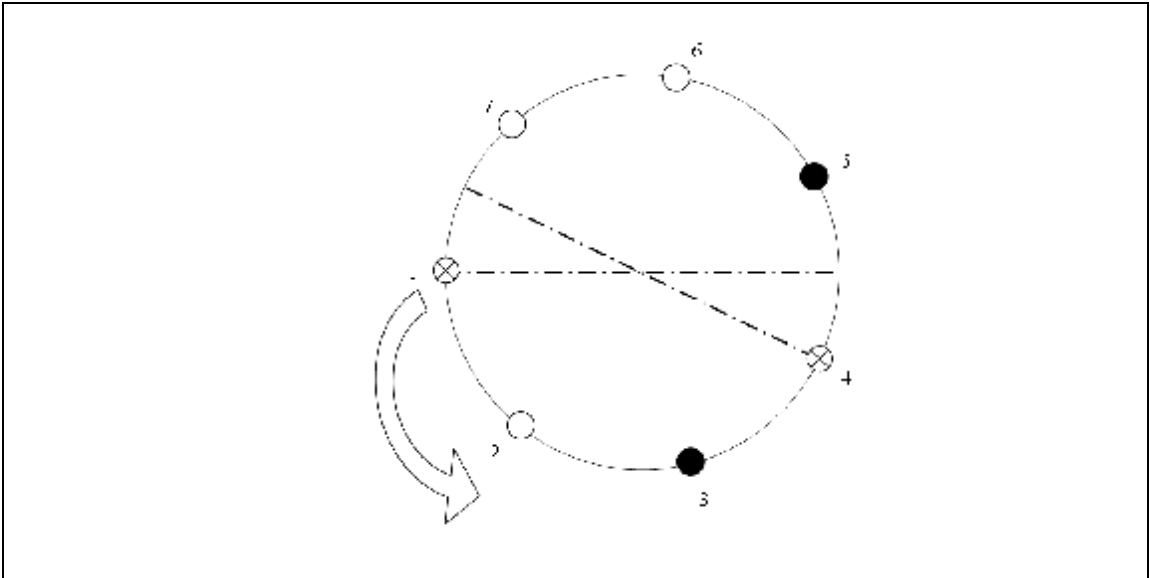

Fig (4): The schematic representation of a ring with an odd number of vertices: As it is seen, each symmetry line of the ring passed through a vertex which is its counterpart. In such a scheme, each intended pair can be considerably entangled only by controlling vertex 1 and a magnetic flux. For example, for targeting vertices 3 and 5, the vertex 4 should be excited. This can be achieved by applying a local magnetic field at vertex 1 and subsequently passing a definite magnetic flux inside the ring to transfer almost completely the excitation to vertex 4.

It is worth mentioning that if we use initial state (19), we can generate nearly perfect entanglement among vertex-pairs that their distance is equal to the distance of the two initialized sites because the applied magnetic flux acts as the transfer operator. In addition, after the transfer, if we permit the system to evolve freely (cutting off the magnetic flux), the efficient provision of W-class *four-partite entanglement* among some selected quadruples of vertices will be feasible since in W-states the excitation-amplitude of qubits should be the same and as high as possible. These states are a resource in some QIP and QC protocols such as telecloning [17], parallel computation [3], and universal error correction [18].

## III. SUMMARY

We considered dynamically interacting spin systems to generate considerable entanglement between on-demand pairs of qubits, given low level of complexity. Due to entanglement monogamy, the sum of entanglements of a vertex with the other vertices can never exceed one e-bit. So, we find that generation of high entanglement between two vertices suffers from entanglement dispersion over all vertex-pairs in generic systems. Accordingly, we searched for structures in which entanglement between some pairs is concentrated. We find out that in axially symmetric structures not only entanglement can be concentrated between equal stratum qubits but also no coding is needed to target them. Moreover, symmetric initialization is as simple as a local excitation if a single element stratum exists. To devise efficient entanglement distributor buses among axially symmetric structures, one-dimensional systems attract our attention due to their bigger achievable entanglement, easier initialization, and simpler implementation. For all pairs to be targetable, uniformly-coupled rings were considered since in such a system, in each pair, two vertices are mirror-picture of one another w.r.t. one of the symmetry lines. To avoid changing the location of the controlling-vertices, we can operate a passing magnetic flux to transfer the excitation from the controlling-vertex to the destination vertex. For example, by using homogeneous rings consisting of 5 or 7 vertices, every pair can be efficiently entangled by exciting one fixed vertex and controlling magnetic flux and time. Due to characteristics listed in section I, such a system can be considered as an efficient entanglement distributor bus, given targetability of every pair, un-engineered couplings, one dimensional structure, and excitation of only one fixed qubit.

Axially symmetric networks could be realized through some well-developed techniques. Recall that a network is "axially symmetric" if the map of interaction couplings is symmetric w.r.t. a line. To establish such a system, it is enough to nest qubits into a spatially axially symmetric graph because the interaction coupling of each qubit pair, which exists due to mutual interaction, depends on the qubits distance. We can create such a system by some fabrication techniques such as 3D-lithography of some types of qubits (e.g. quantum dots, or superconducting qubits) because we are able, at least in theory, to fabricate each qubit in the desired spatial position. Our final proposed system can also be manufactured by 2D-lithography where a nanometric coil is also fabricated at the ring's center.

These special structures are highly interesting for application in nanoscale quantum information chips because they can simultaneously serve as qubits, information transmission lines, and according to the present work, as efficient entanglement distribution buses by only one simple contact point with the external environment since, in such scales, minimization of employed resources for fulfilling different necessary parts is a critical issue.

**Acknowledgements:** Special thanks to Prof Shahin Rouhani, Hamid Afshar, Mohammad Vahedi and Meysam Rahimi for careful reading of the manuscript and their helpful suggestions.

**Appendix A: Calculation of entanglement between vertex-pairs**

In order to compute the entanglement generated between two qubits $m$ and $n$, we must determine the reduced density matrix of these two qubits. For calculation of the related density matrix, the system density matrix $|\Psi(t)\rangle\langle\Psi(t)|$ must be traced over all vertices except $m$ and $n$. In the case of (6), after some calculation, the result is

$$\rho_{mn}(t) = \begin{pmatrix} 1-Cos^2(\theta)(|A_m|^2+|A_n|^2) & \frac{1}{2}Sin(2\theta)e^{-i\varphi}A_n & \frac{1}{2}Sin(2\theta)e^{-i\varphi}A_m & 0 \\ \frac{1}{2}Sin(2\theta)e^{i\varphi}A_n^* & Cos^2(\theta)|A_n|^2 & Cos^2(\theta)A_m A_n^* & 0 \\ \frac{1}{2}Sin(2\theta)e^{i\varphi}A_m^* & Cos^2(\theta)A_n A_m^* & Cos^2(\theta)|A_m|^2 & 0 \\ 0 & 0 & 0 & 0 \end{pmatrix}$$

In order to calculate the concurrence of $\rho_{mn}(t)$ [13], the matrix $R$ must be determined

$$\sigma_y \otimes \sigma_y = \begin{pmatrix} 0 & 0 & 0 & -1 \\ 0 & 0 & 1 & 0 \\ 0 & 1 & 0 & 0 \\ -1 & 0 & 0 & 0 \end{pmatrix} \rightarrow$$

$$R = \begin{pmatrix} 0 & Sin2\theta Cos^2\theta e^{-i\phi}|A_i|^2 A_j & Sin2\theta Cos^2\theta e^{-i\phi}|A_j|^2 A_i & \frac{-1}{2}Sin^2 2\theta e^{-2i\phi} A_i A_j \\ 0 & 2Cos^4\theta|A_i|^2|A_j|^2 & 2Cos^4\theta|A_j|^2 A_i A_j^* & -Sin2\theta Cos^2\theta e^{-i\phi}|A_j|^2 A_i \\ 0 & 2Cos^4\theta|A_i|^2 A_i^* A_j & 2Cos^4\theta|A_i|^2|A_j|^2 & -Sin2\theta Cos^2\theta e^{-i\phi}|A_i|^2 A_j \\ 0 & 0 & 0 & 0 \end{pmatrix}$$

The eigenvalues of $R$ are the roots of $\det(R-\lambda\mathbf{1})=0$ and we find

$$(-\lambda)^2\left[(2Cos^4\theta|A_i|^2|A_j|^2-\lambda)^2 - 4Cos^8\theta|A_i|^4|A_j|^4\right]=0 \rightarrow \lambda_1=\lambda_2=0$$

$$(2Cos^4\theta|A_i|^2|A_j|^2-\lambda)^2 - 4Cos^8\theta|A_i|^4|A_j|^4 = 0 \rightarrow \lambda_\pm = 2Cos^4\theta|A_i|^2|A_j|^2(1\pm 1) \rightarrow$$
$$\lambda_-=0, \lambda_+=4Cos^4\theta|A_i|^2|A_j|^2$$

So in descending order, the square root of eigenvalues are $\left(2Cos^2\theta|A_i A_j|;0;0;0\right)$ and the entanglement between vertices $i$ and $j$ is obtained as

$$C_{ij} = \max\left(0, 2Cos^2\theta|A_i A_j|-0-0-0\right) = 2Cos^2\theta|A_i(t)A_j(t)|$$

**Appendix B: the Hamiltonian of a uniformly-coupled XXZ ring in the presence of a magnetic flux and its diagonalization**

In this case, the hopping term of Hamiltonian acquires a pure phase $e^{i\Phi}$ [19] i.e.

$$H = -\frac{J}{2}\sum_{k=1}^{N}\left[(e^{i\Phi}\sigma_k^+.\sigma_{k+1}^- + H.c.)\right] + \Delta\sum_{k=1}^{N}\sigma_k^z.\sigma_{k+1}^z$$

Where $\Phi$ is the magnetic flux. This Hamiltonian can be mapped to a spinless fermion model by Jordan-Wigner transformations

$$H = -\frac{J}{2}\sum_{k=1}^{N}\left[(e^{i\Phi}c_k^\dagger c_{k+1} + H.c.)\right] + J\Delta\sum_{k=1}^{N}(c_k^\dagger c_k - \frac{1}{2})(c_{k+1}^\dagger c_{k+1} - \frac{1}{2})$$

For rotating fermions in a closed path, the flux can be gauged out of the Hamiltonian so that solving the Schrödinger equation in the presence of the magnetic flux with a periodic boundary condition is equivalent to that in the absence of the magnetic flux but with a twisted boundary condition for the wave-functions [20]

$$\Psi(x_1,...,x_i,...) = e^{i\Phi}\Psi(x_1,...,x_i+L,...)$$

Under this boundary condition, time independent Schrödinger equations can be solved by Bethe ansatz [21].